\documentclass{jfm}

\usepackage[latin1]{inputenc}
\usepackage{graphicx}
\usepackage{amsmath}
\usepackage{natbib}
\usepackage{bm}

\usepackage[active]{srcltx}
\usepackage{color}

\newcommand{\beq}{\begin{equation}}
\newcommand{\eeq}{\end{equation}}
\newcommand{\ba}{\begin{eqnarray}}
\newcommand{\ea}{\end{eqnarray}}
\newcommand{\bee}{\begin{eqnarray}}
\newcommand{\eee}{\end{eqnarray}}
\newcommand{\bc}{\begin{center}}
\newcommand{\ec}{\end{center}}

\begin{document}

\title[Dynamics of elastic dumbbells sedimenting in a viscous fluid]{Dynamics of elastic dumbbells sedimenting in a viscous fluid: oscillations and hydrodynamic repulsion}
\author[M. Bukowicki, M. Gruca, M. L. Ekiel-Je\.zewska]{Marek
Bukowicki$^1$, Marta Gruca$^1$\break and Maria L. Ekiel-Je\.zewska$^1$
\thanks{Email address for correspondence: mekiel@ippt.pan.pl},\ns}
\affiliation{$^1$Institute of Fundamental Technological Research,
Polish Academy of Sciences, \\02-106 Warsaw, Pawi\'nskiego 5$b$, Poland\\[\affilskip]}
\date{\today}
\maketitle
\begin{abstract}
Hydrodynamic interactions between two identical elastic
dumbbells settling
under gravity in a viscous fluid at low-Reynolds-number
are investigated within the point-particle model. Evolution
of a benchmark initial configuration is studied, in which
the dumbbells are vertical and their centres are aligned
horizontally. Rigid dumbbells and pairs of separate beads
starting from the same positions tumble periodically while
settling down. We find that elasticity (which
breaks time-reversal symmetry of the motion) significantly affects the
system's dynamics. This is remarkable taking into account that elastic forces are always much smaller than gravity. We observe oscillating motion of the
elastic dumbbells, which tumble and change their length
non-periodically. Independently of the value of the spring
constant, a horizontal hydrodynamic repulsion appears between
the dumbbells - their centres of mass move apart from each other
horizontally. The shift is fast for moderate values of the spring
constant $k$, and slows down when $k$ tends to zero or to infinity;
in these limiting cases we recover the periodic dynamics reported
in the literature. For moderate values of the spring constant, and
different initial configurations, we observe the existence of a
universal time-dependent solution to which the system converges
after an initial relaxation phase. The tumbling time and the width
of the trajectories in the centre-of-mass frame increase 
with time. 
In addition to its fundamental significance, the benchmark
solution presented here is important to understand general features of
systems with larger number of elastic particles, at regular and random
configurations.
\end{abstract}

\begin{keywords}
Stokesian dynamics, hydrodynamic interactions, elastic dumbbells, periodic
sedimentation
\end{keywords}

\section{Introduction}\label{sec:intro}

Dynamics of elastic particles, both sedimenting and entrained by an ambient
flow, attracts a lot of attention owing to importance of many-particle
systems of flexible fibres, vesicles or capsules
for industry, biology, medicine   and nano-technology, as emphasized e.g.
by \cite{KantslerGoldstein}, \cite{Vlahovska},  \cite{Dreyfus},
\cite{Drescher}. Deformation and the existence of constraint forces  leads
to a very complex dynamics even in case of isolated particles, and to
significant modifications of hydrodynamic interactions between such
objects. For example, there appear different modes of the dynamics, analysed e.g. by
\cite{YoungShelley}, or migration and accumulation of isolated flexible
particles in ambient flows, as shown e.g. by  \cite{Aga,Farutin}.

For very elongated particles, the effects of bending are of course significant, and the 
worm-like chain approximation is usually applied, e.g. by \cite{YoungShelley,KantslerGoldstein,Aga}. 
However, for confined polymers, the bead-spring model and a simple Hookean dumbbell are often 
used to reproduce the essential features of their dynamics in the simplest way, see e.g. \cite{Jendrejack,Shaqfeh,Graham}. Each dumbbell consists of two identical spherical beads, 
with their centers connected by infinitely thin spring, which does not interact with the fluid flow. 
Evolution of elastic dumbbells in various ambient flows have recently been determined as useful 
simple benchmarks - in analogy to the classical solutions for rigid dumbbells, outlined e.g. in 
the classical review of \cite{Brenner}. 
It is known that for non-deformable particles
entrained by an ambient flow or sedimenting under gravity, there appear
oscillating solutions, e.g. those constructed by \cite{Graham} or \cite{HolzerZimmermann},
important for the dynamics which starts from a wide range of the  initial
configurations.
In many cases, sedimenting particles which perform quasi-periodic
oscillations or complex chaotic Stokesian dynamics seem to be close to
periodic solutions, as proposed by \cite{Janosi} and further analysed  by
\cite{Ekiel-Jezewska2}.
Periodic motions of sedimenting particles (\cite{Hocking},
\cite{Jayaweera}, \cite{Caflisch}) have fundamental
significance for deeper understanding the complex dynamics of many-particle
systems (\cite{Ekiel-Jezewska}).

An interesting question is to what extent the dynamics of elastic particles
is influenced by the existence (or lack) of
periodic solutions similar to those observed for the rigid particles.
For flexible objects in shear or Poiseuille flows, it was shown that
elasticity often prevents the occurrence of periodic motion. Typically,
owing to the lack of the time-reversal symmetry,
there appear a migration  of the orbits analysed by \cite{Jeffery}, as discussed e.g. by \cite{Graham}.

The effect of elasticity on systems of sedimenting particles has not yet
been systematically investigated. Therefore, the goal of this paper is to
determine the influence of elastic constraints on the relative motion of a
benchmark system of two heavy 
dumbbells. We 
focus on the class of the regular initial conditions  which
was studied before
by \cite{Jung} and \cite{Hocking}.
There exists a close analogy between periodic relative motion of  four
heavy point-like  beads, found theoretically by \cite{Hocking},  and
periodic sedimentation of
a pair of elongated rigid particles, observed experimentally and
numerically by \cite{Jung}.
Tracking the ends of elongated particles it turns out that their periodic
relative trajectories are qualitatively the same as the trajectories of
four point-like beads, providing that initially they are located at
vertices of a rectangular with vertical sides determined by the
corresponding  positions of the rigid particles.
Therefore, a question arises how the dynamics looks like in case of two
{\it elastic} dumbbells, initially at the same positions
as the rigid ones. Do periodic solutions also exist?
In this work, we characterize the main features of the dynamics of elastic
dumbbells at such benchmark initial configurations.

\section{Model}\label{sec:model}

We consider two identical elastic dumbbells falling under gravity in a fluid of viscosity $\eta$, at low-Reynolds-number. Each 
dumbbell consists of two identical heavy beads of radius $a$ connected by a spring with a spring constant $k$ and equilibrium length $L_0$. Therefore, 
the external force $\mathbf{F}_i$ acting on each bead \textit{i} is the sum of the gravitational force
$\mathbf{G} = (0,0,-G)$ (with $G>0$) and the elastic force $\mathbf{S}_i$, 
\bee
\mathbf{F}_i = \mathbf{G} + \mathbf{S}_i,\label{Fext}
\eee
where
\bee
\mathbf{S}_1 &=& - \mathbf{S}_2 = - S \hat{\bm{r}}_{12},\\
S& =& k (r_{12}-L_0),\label{S1}
\eee
with the position of bead $i$ denoted as ${\bm r}_{i}$, and
 ${\bm r}_{ij}={\bm r}_i-{\bm r}_j$, $r_{ij}=|{\bm r}_{ij}|$, $\hat{\bm r}_{ij}={\bm r}_{ij}/r_{ij}$. Analogical expressions for the forces 
$\mathbf{S}_3$ and $\mathbf{S}_4$ are obtained from Eq.~\eqref{S1} by the replacements $1\rightarrow 3$ and $2 \rightarrow 4$. 

\begin{figure}
\centerline{\includegraphics[height=4.8cm]{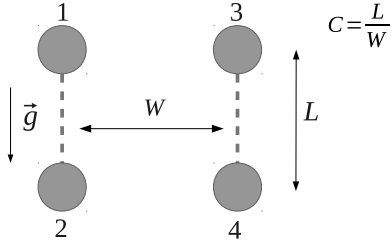}}
\vspace{-.8cm}
\caption{A characteristic ``vertical'' configuration of dumbbells considered in this work.}
\label{fig:schemat_init}
\end{figure}

Initially, the dumbbells are oriented vertically, with their centres aligned horizontally, as illustrated in Fig.~\ref{fig:schemat_init}. 
The distance $L_{in}\equiv L(t=0)$ between the beads within dumbbells is equal to the spring equilibrium length, $L_0$, or $1.7L_0$ and the 
distance $W_{in}\equiv W(t=0)$ between the centres of dumbbells spans a range of values determined by the configuration aspect-ratio
 $0.9 \le C_{in}\equiv L_{in}/W_{in}\le 1.8$.

We assume that the Reynolds number is much smaller than unity. In this case, the bead velocity relaxation time $\tau=m/\zeta$, estimated as the bead mass $m$ divided by its friction coefficient $\zeta=6\pi\eta a$, is much smaller than the hydrodynamic time scale, determined by the ratio of the bead radius and its velocity; the inertia is irrelevant. The total force (the sum of the hydrodynamic and the external
forces) exerted on each bead vanishes. Therefore, the bead velocities are determined by their positions. 

In the low-Reynolds-number regime, the fluid velocity $\bm{v}(\bm{r})$ and pressure $p(\bm{r})$ satisfy the Stokes equations. In general, they are solved with specific boundary conditions imposed on the bead surfaces (typically, it is assumed that the fluid sticks to the solid surfaces of the beads) and at the system boundaries - in our case, there is no ambient flow, the fluid is unbounded and motionless at infinity. The resulting motion of the beads is sometimes called the Stokesian dynamics. It is often determined by the multipole expansion (see \cite{Felderhof}), which can be interpreted as multiple `reflections' (see  \cite{KimKarrila}) or `scatterings' (see \cite{Felderhof}) of the flow by the bead surfaces. 
We use the standard dumbbell model, in which the spring is attached to the bead centers. Within this model, there is no external torques exerted on the particles. Therefore, according to the classical mobility problem based on the Stokes equations, see e.g. \cite{KimKarrila} or \cite{Felderhof}, time derivatives of the bead positions, equal to components of their translational velocities, are sums of the corresponding translational-translational mobility coefficients multiplied by the components of the external forces $\mathbf{F}_i$ exerted on each bead $i$, see Eqs. \eqref{Fext}-\eqref{S1}. The mobility coefficients and the external forces depend on the bead relative positions, but do not depend on their orientations. Owing to this property, it is possible to 
determine translational motion of the beads by solving the system of ODEs for their time-dependent positions. This can be done 
without knowing anything about time-dependent particle orientations - translational motion is decoupled from the rotational one.

In this work, the bead radius $a$ is assumed to be much smaller than the distances between the bead centres, and therefore the point-force approximation is justified.
 In this case, 
see e.g. \cite{KimKarrila}, the fluid velocity $\bm{v}(\bm{r})$ and pressure $p(\bm{r})$ satisfy the Stokes equations with the additional sum of the point forces, located at the centres of the beads $\bm{r_i}$, $i=1,...,4$,

\begin{eqnarray}
\label{stokes_eq}
	\eta \nabla^2 \bm{v(r)} - \nabla p(\bm{r}) = -\sum_{i = 1}^{4}\mathbf{F}_i\delta(\bm{r - r_i}), \nonumber \\
\nabla \cdot \bm{v(r)} = 0.
\end{eqnarray}
The only additional condition is that the fluid velocity vanishes at infinity (there are no requirements at the bead surfaces). Then,  $\bm{v}(\bm{r})$ and $p(\bm{r})$ follow as superpositions of the corresponding Green tensors ${\bm G}(\bm{r}-\bm{r_i})$, contracted with the forces $\mathbf{F}_i$, $i=1,...,4$. For velocity, ${\bm G}$ is the Oseen tensor
$\mathbf{T}(\bm{r})$,  
\begin{equation}
	\mathbf{T(r)} = \frac{1}{|\bm{r}|}({\bm I} + \frac{\bm{r} \otimes \bm{r}}{|\bm{r}|^2}).
\end{equation}

Here and later on, we use dimensionless variables, based on $L_0$ as the length unit, 
${G}/(8 \pi \eta L_0)$ as the velocity unit, and $G$ as the force unit. Therefore, 
 ${8 \pi \eta L_{0}^2}/G$ is the time unit, and ${G}/{L_{0}}$ is the unit of the spring constant.

In the point-particle model, each bead is  advected by the superposition of fluid flows generated by all the other point-like beads, plus a self term, proportional to the single-particle mobility and the external force. Therefore, velocity  ${\bm V}_i$ of a particle $i$ has the form,
\begin{equation}
 {\bm V}_i = \frac{4}{3a} \mathbf{F}_i + \sum_{j\neq i} \mathbf{T}(\bm{r}_{ij}) \cdot \mathbf{F}_j,
  \label{pp}
\end{equation}

Positions $\bm{r}_i$ of the beads $i = 1,\ldots, 4$ as functions of time are determined by the set of ODEs given below,
\bee
 \frac{d\bm{r}_i}{dt} = {\bm V}_i.\label{eqm}
\eee

In Eq. \eqref{pp}, the bead radius enters only into the self-term, through the single-bead mobility coefficient. In the absence of elasticity, a frame of reference moving with a single bead can be chosen, i.e. with velocity $G/(6\pi\eta a)$. In this frame, the dynamics does not depend on the bead radius, see \cite{Ekiel-Jezewska}. It means that the relative motion does not depend on the bead radius. However, in the presence of the time-dependent spring forces ${\bf S}_i$, such a reasoning cannot be applied. Taking into account that $S<< G$ (as it will be explained in Section \ref{subsec:comparison}) we conclude that in the presence of elastic forces, the relative motion of the beads depends on $a$, but this effect is very small. 

The equations of motion \eqref{pp}-\eqref{eqm} are solved numerically
using the 4th order adaptive Runge - Kutta method.
 
\section{Results}\label{sec:results} 

\subsection{Basic features of the dynamics\label{sec:basic}}
Assume that initially four beads are placed at the corners of a vertical rectangle as shown in Fig.~\ref{fig:schemat_init}, and consider pairs of the beads
 1-2 and 3-4 as ``dumbbells''. Let's first discuss the dynamics for $k = 0$, i.e. for separate beads without any constraint forces, analysed 
by \cite{Hocking,Tory} and shown as videos by \cite{Sikora}, and for $k = \infty$, i.e. for two rigid dumbbells, which model rods, investigated
 by \cite{Jung}. At the initial configuration hydrodynamic interactions between the beads lead to contraction of the upper side of the rectangle 
and elongation of the bottom one, so an isosceles trapezoid  is formed. Later on, the upper side expands a bit, while the bottom side expands even more. 
The beads from the upper side move faster than the others, the height of the trapezoid monotonically diminishes and the dumbbells rotate. At the same time,
 hydrodynamic interactions lead to drifting apart of the right and left dumbbells and the trapezoid expands in the horizontal direction. 
For $k=0$, the dumbbell length $L$ shrinks, while for $k=\infty$, it is constant. After a certain time $\tau$, the trapezoid flattens to a single horizontal
 line and the largest horizontal distance between the beads reaches the maximum. 

\begin{figure}
\hspace{-0.2cm}\centerline{\includegraphics[width=10.3cm]{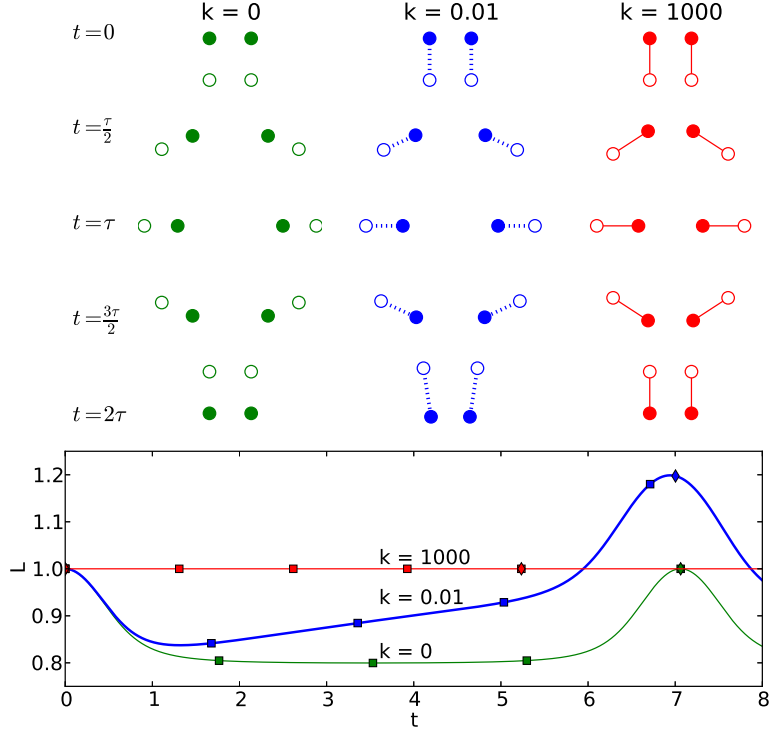}}\vspace{-0.3cm}
\caption{Evolution of pairs of separate beads (k=0, green online), elastic dumbbells (k=0.01, blue online) and almost rigid dumbbells (k=1000, red online), for $a\!=\!0.1$, $C_{in}\!=\!1.0$ and $L_{in}\!=\!1.0$. Top: snapshots taken at equal time intervals $\tau/2$. %
Bottom: evolution of the dumbbell length $L$, with squares corresponding to the snapshots, and diamonds to
vertical configurations. For $k=0.01$, the motion is not periodic.
}
\label{fig:evo}
\end{figure}

For larger times, $\tau +t$, the particle positions coincide with the earlier positions of the particles at time $\tau - t$, reflected in this horizontal plane in which all the particles are located at time $\tau$. This property follows from the symmetry of the Stokes equations with respect to the time reversal, as explained e.g. by \cite{Caflisch}.
Therefore, the internal beads move faster than the external ones and the upside-down trapezoid is formed. The left and right dumbbells rotate and approach
each other, and after the tumbling time $2\tau$ they again form the initial rectangle, but with the flipped dumbbells -  now the beads 2 and 4 are at the top, see the first and last plots of the upper panel in Fig.~\ref{fig:evo}.

\begin{figure}
  \centerline{\includegraphics[height=4.4cm]{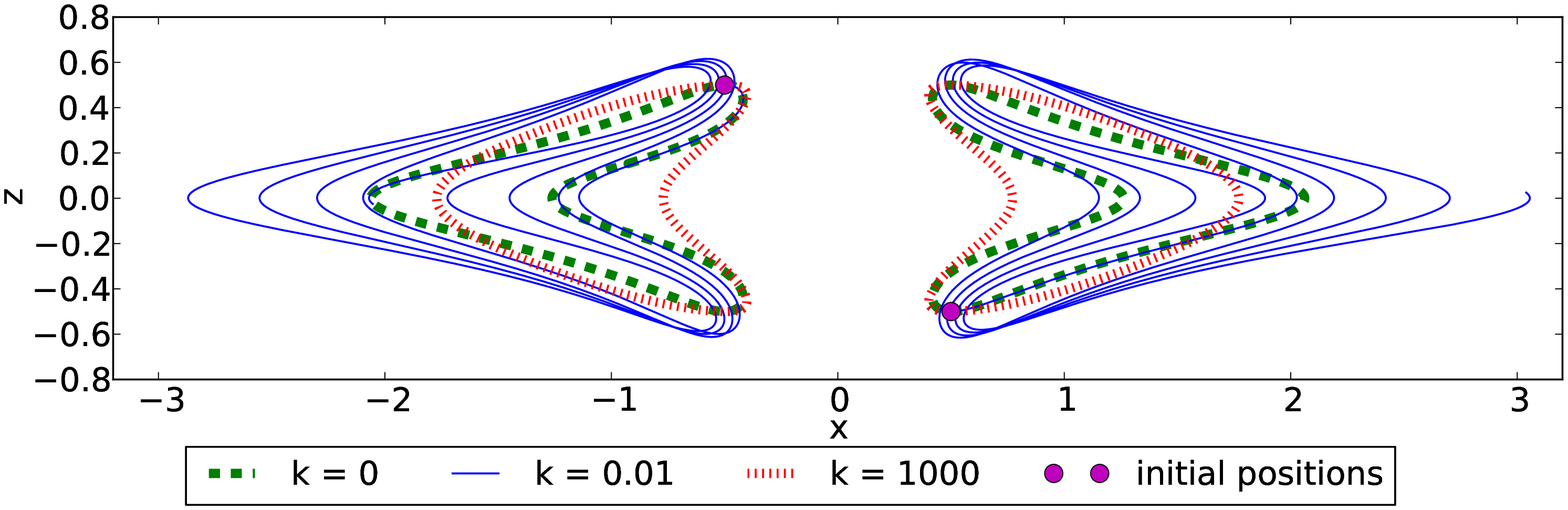}}\vspace{-0.1cm}
  \caption{Trajectories of two beads from different dumbbells, in the  centre-of-mass reference frame (drawn to scale). They are closed (periodic) for separate beads and rigid dumbbells, but non-closed (non-periodic) for elastic dumbbells. 
Here, $a\!=\!0.1$, $C_{in}\!=\!1$, $L_{in}\!=\!1$.}
\label{fig:trajec}
\end{figure}

Basic features of this tumbling dynamics are recovered also by sedimenting elastic dumbbells as they move from a vertical configuration to a horizontal one
and vice versa. 
The essential difference is that, owing to the presence of the elastic 
forces \eqref{S1}, there is no symmetry between the relative motion of the beads at $\tau -t$ and $\tau + t$. As the result, there is no periodicity of the 
relative positions, and consecutive tumbling times are not equal to each other. In Fig.~\ref{fig:evo}, the lack of symmetry is evident in the middle plot of the top panel, where the initial vertical dumbbell configuration at $t=0$ is not reproduced at $t=2\tau$.
It is also
clearly visible in the bottom panel as the non-symmetric blue curve of the time-dependent dumbbell length $L$. In particular, the second vertical 
configuration (marked by blue diamond) corresponds to time larger than $2\tau$ (marked by blue square). For comparison, we also show the green symmetric 
curve for separate beads, and the red almost constant curve for $k=1000$, to illustrate that such a dumbbell can be practically considered as the rigid 
one - indeed, in this case $L$ changes by $3 \cdot 10^{-5}$ only.

It is convenient to analyse the influence of elasticity on the motion, using the centre-of-mass reference frame, so periodicity can be easily visible
 as closed trajectories. In Fig.~\ref{fig:trajec}, we plot the centre-of-mass trajectories of two beads from different dumbbells.
For $k = 1000$ (red dotted line) the trajectories, plotted for 16 tumbling times, are practically closed, and their shape is qualitatively similar
 (but not identical) to the closed trajectories of the periodic motion of separate beads ($k = 0$, green dashed line). 
Beads of the elastic  dumbbells ($k = 0.01$, blue solid line) perform a non-periodic motion and their trajectories are not closed. Until the second tumbling, the trajectories
stay close to their analogues for $k=0$ and $k=1000$, but later on they systematically deviate from the periodic shapes and become wider, as shown in Fig. \ref{fig:trajec}. 
The widening of trajectories has been observed during the whole time of simulations and for all intermediate values of $k$, so it seems to be an 
intrinsic property of the studied system. The widening is correlated with the increase of the trajectory length between subsequent tumbling times, 
increase of the tumbling time, and increase of the maximal horizontal distance between the dumbbells, reached at their horizontal orientations.

\subsection{Relaxation to a universal trajectory}
\begin{figure}
  \centerline{\includegraphics[width=13.7cm]{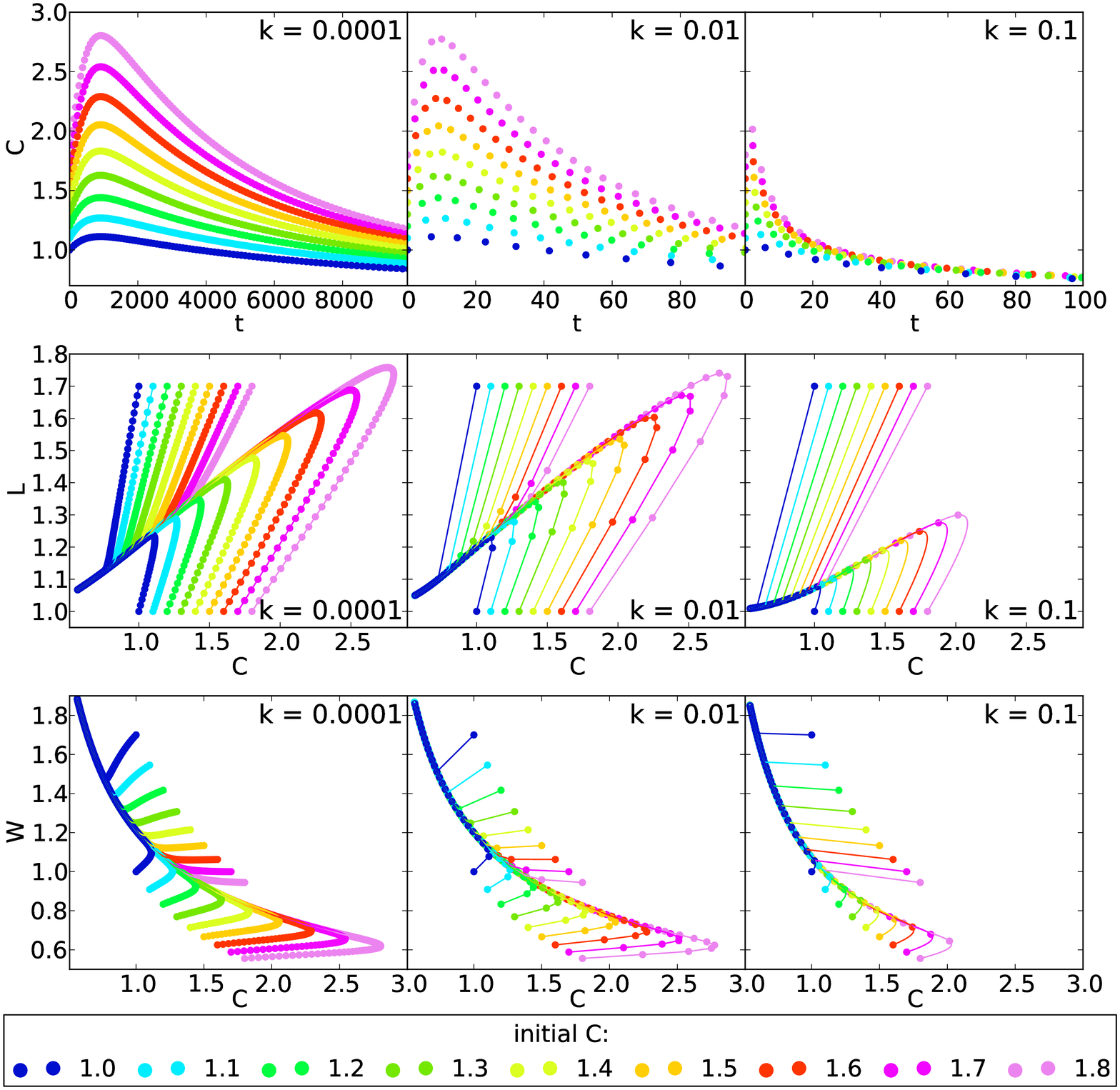}}
\vspace{-0.4cm}
  \caption{The system relaxation to a universal time-dependent solution. 
The dots represent consecutive vertical configurations with time-dependent aspect ratio $C$, height $L$ and width $W$. Top: evolution of $C$ (for $L_{in}=1$). 
Middle: $L$ versus $C$; bottom: $W$ versus $C$ (for $L_{in}$=1 or 
1.7). 
 }
\label{fig:ct}
\end{figure}

As the motion of elastic dumbbells is not periodic, the parameters $C$, $L$ and $W$, measured in vertical configurations 
(shown in Fig.~\ref{fig:schemat_init}), change over time and can be used to describe evolution of the dynamics. The results are shown in Fig.~\ref{fig:ct}. 
Here, the simulation time $t=10000$ and $a=0.1$. Two main phases of the dynamics are clearly visible. For a given $k$, the system converges to a single universal time-dependent solution.

In the top panel, the time-dependent aspect ratio $C$ of various vertical configurations with $L_{in}=1$ is shown. 
During the first phase, the value of $C$ raises. The relaxation occurs, since the state when vertical dumbbells keep their equilibrium length $L_0$ does
 not seem to be ``natural'' for the studied system. As illustrated in the bottom panel of Fig.~\ref{fig:evo}, elastic dumbbells tend to be stretched at
 vertical orientations. In the second phase, value of $C$ monotonically decreases and approaches a $k$-specific limit, $C_0$. The stiffer the dumbbells 
are, the relaxation time becomes shorter, as shown in the top panel of Fig.~\ref{fig:ct} (note different time scales displayed at different values of $k$). 
This behaviour is observed for $0.0001\le k \le 1$. 

To investigate the dynamics after the relaxation time, 
in the lower part of Fig.~\ref{fig:ct} we plot $L$ versus $C$ and $W$ versus $C$, measured at consecutive vertical 
configurations. The essential feature following from these plots is that all the initial vertical configurations converge to a single, universal trajectory. 

Independently of the initial values  $L_{in}$ of the dumbbell length, and $C_{in}$ of the system aspect ratio, after a certain relaxation time, $C$ monotonically decreases in consecutive vertical configurations. Then, evolution
 of the trajectories $W(C)$ and $L(C)$ is the same and depends only on the spring constant $k > 0$. The same vertical configurations lead to identical,
 universal time-dependent evolution in-between. Relaxation of the system, clearly visible in the middle and bottom panels of Fig.~\ref{fig:ct}, takes place after a time scale which
 can be estimated from the top panel of the same figure. 

\subsection{Hydrodynamic repulsion}\label{sec:repulsion}
\begin{figure}
  \centerline{\includegraphics[width=7.8cm]{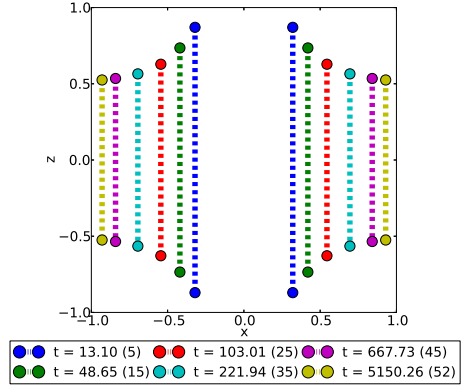}}\vspace{-0.2cm}
  \caption{The snapshots of elastic dumbbells at subsequent vertical configurations, after the indicated time and number of flips. The dumbbells move away 
from each other during the motion. Here, $k \!= \!0.01$, $a\!=\!0.1$, $C_{in}\!=\!1.8$ and $L_{in}\!=\!1.0$. For the sake of figure clarity, the bead size
 is reduced three times.}
  \label{fig:hyd_rep}
\end{figure}

From Fig.~\ref{fig:ct} it follows that in consecutive vertical configurations of the universal trajectory, the distance $W$ between the dumbbells 
increases with time. This behaviour can be interpreted as an effective horizontal repulsion of elastic dumbbells, caused by their hydrodynamic interactions.
The effect is clearly visible in Fig. \ref{fig:hyd_rep}, where relative positions of the beads are shown in subsequent (but not consecutive) vertical 
configurations. Systematically, for a larger time, the distance between vertically oriented dumbbells is larger. 
The hydrodynamic repulsion can be significant; e.g., as illustrated in Fig.~\ref{fig:ct}, it can lead to a triple increase of the distance $W$ between 
the dumbbells at vertical configurations. The speed of the repulsion is the largest for $k=0.01-0.1$.

More information about hydrodynamic repulsion and other features of the dynamics of two sedimenting elastic dumbbells can be found  in two supplementary
movies, which illustrate the motion in the centre of mass reference frame. Evolution of two initial aspect ratios, $C_{in}\! = \!1$ and $C_{in}\!=\!1.8$ 
is shown for two values of the spring constant, $k\!=\!0.01$ and $k\!=\!0.1$. 
Initially, the dumbbells have the equilibrium length $L_{in}\!=\!1$, but it is clearly visible that their length $L$ changes  with time while the dumbbells
 tumble. This effect is pronounced for $C_{in}\! = \!1.8$ and $k\!=\!0.01$. 

At the beginning, the dumbbell length $L$ at consecutive vertical positions increases with time, what corresponds to the relaxation phase, in agreement
with  Fig.~\ref{fig:ct}. Later, when the universal trajectory is approached, the value of $L$ at consecutive vertical positions decreases monotonically.
 For $C_{in}\!=\!1.8$, the initial horizontal distance $W_1$ between vertical dumbbells is much smaller than $W_2$ in case of $C_{in}\! =\! 1$. Therefore,
 the universal trajectory is approached at a smaller value of $W_1$, and more time is needed for the system to reach the larger value $W_2$. Comparing
 the movies for $C_{in}\!=\!1.8$ and $C_{in}\! =\! 1$, we can observe that at the universal trajectory, the tumbling time significantly increases when
 $W$ evolves and becomes larger. 
At our movies, the largest speed of the hydrodynamic repulsion is observed for $k=0.1$ and $C_{in}=1.8$. This effect results from an interplay of the 
the corresponding slopes of the $W(C)$ and $C(t)$ curves in the relevant regimes. 

At the universal trajectory, the parameter $C$ decreases with time, what leads to universal patterns of the trajectory widening, including the increase 
of its length and the maximal horizontal size $d_{\text{max}}$.  
The analogous dependence of the trajectory width and the tumbling time on the parameter $C$ was observed by \cite{Ekiel-Jezewska} for a wider 
class of regular systems of separate beads.

\subsection{How large are elastic forces in comparison to gravity?}\label{subsec:comparison}

The significant hydrodynamic repulsion effect shown in Fig. \ref{fig:hyd_rep} 
is entirely due to elastic forces between the beads which form a dumbbell (it is absent in case of rigid constraints or separate beads). Therefore, in this section we investigate how large are the elastic forces $S$ in comparison to gravity $G$. As shown in Eq.~\eqref{S1}, $S$ is proportional to the deviation of the spring length from its equilibrium value, $r_{12}-L_0$. This quantity depends on the system dynamics, i.e.  evolution of all the bead positions. While evolving, the system adjusts these positions in a way which depends on all the system parameters (in particular, also on the bead radius).

It is remarkable that the large hydrodynamic repulsion is observed even though the elastic force is 
an order of magnitude lower than gravity, as we discuss below. The following arguments are valid in case of the bead radius $a \ll r$, where $r$ is 
the distance between beads. The velocity of  bead 1 is given by:
\begin{equation}
	\bm{V}_1 = \frac{4}{3a}\bm{S}_1 + \bm{T}(\bm{r}_{13})\cdot(\bm{G} + \bm{S}_3) + \bm{T}(\bm{r}_{14})\cdot(\bm{G} + \bm{S}_4) + \bm{T}(\bm{r}_{12})\cdot(\bm{G} + \bm{S}_2)\label{for}
\end{equation}

The velocity of  bead 2 of the same dumbbell is easily obtained from Eq.~\eqref{for} by the replacements: $1 \rightarrow 2, \quad 2 \rightarrow 1$.
The maximum value of elastic force $S= k(L - L_0)$ is reached when $dL/dt=0$, i.e. when the relative translational velocity of the beads of each dumbbell has
 the vanishing component parallel to the dumbbell. (In general, the maximum does not occur at vertical configurations.) After rearrangements we obtain,
\begin{eqnarray}
\begin{split}
\hat{\bm{r}}_{12} \cdot (\bm{V}_2 - \bm{V}_1) = \hat{\bm{r}}_{12} \cdot \left[ S \frac{-8}{3a}\hat{\bm{r}}_{12}
- G \Bigl(\bm{T}(\bm{r}_{23})+ \bm{T}(\bm{r}_{24}) -\bm{T}(\bm{r}_{13}) - \bm{T}(\bm{r}_{14})\Bigl)\cdot \hat{\bm{z}}\right.\nonumber \\ \left.
+ S \Bigl(2 \bm{T}(\bm{r}_{12})\cdot\hat{\bm{r}}_{12} + \bm{T}(\bm{r}_{23})\cdot\hat{\bm{r}}_{34} - \bm{T}(\bm{r}_{24}) \cdot\hat{\bm{r}}_{34}  - \bm{T}(\bm{r}_{13})\cdot \hat{\bm{r}}_{34} + \bm{T}(\bm{r}_{14}) \cdot\hat{\bm{r}}_{34}\Bigl) \right] = 0
\end{split}\\
\label{eq:dlugie}
\end{eqnarray}

\begin{figure}
 \centerline{
\includegraphics[width=11cm]{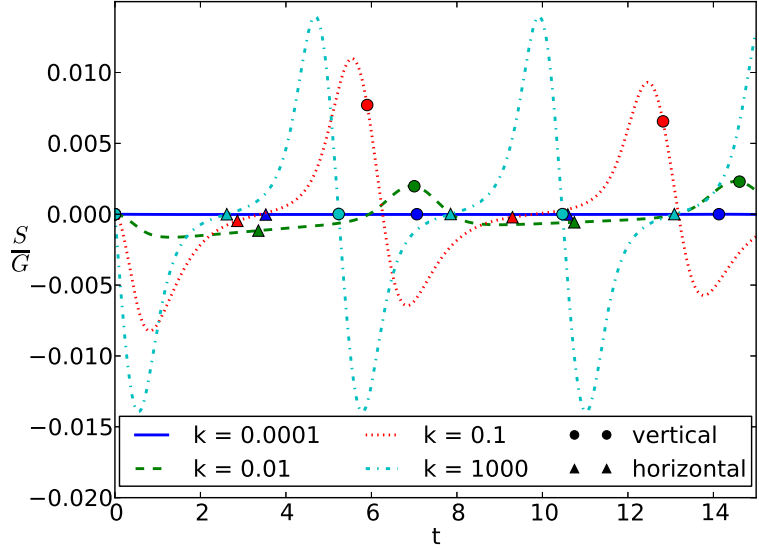}}
  \caption{The ratio $\frac{S}{G}$ of the elastic to gravitational forces is much smaller than one. Symbols: vertical and horizontal dumbbell configurations.}
\label{fig:forces}
\end{figure}
\begin{figure}
  \centerline{\includegraphics[width=14cm]{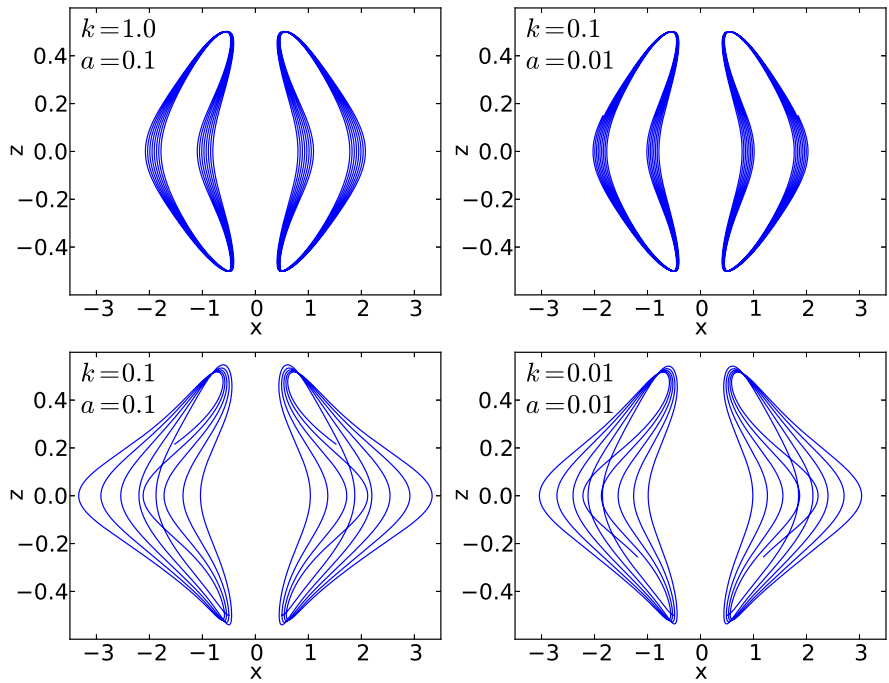}}\vspace{-0.5cm}
  \caption{Trajectories of dumbbell beads in the centre-of-mass reference frame are practically determined only by the ratio $k/a$, providing that $a<<1$. 
Top: $\frac{k}{a} = 10$. Bottom: $\frac{k}{a} = 1.0$. }
\label{fig:trajectories_fig}
\end{figure}

At the r.h.s. of Eq. \eqref{eq:dlugie}, the first and the third terms are proportional to the elastic forces $S$. The first term scales as
 $ \sim \frac{1}{a}$ and the following ones scale as  $T(\bm{r}) \sim \frac{1}{r}$. Therefore, for $a \ll r$, the third term can be neglected.
Given that the relative translational velocity of the beads along the dumbbell vanishes, we obtain
$\frac{|{S}|}{a} \sim \frac{\bm{G}}{r} \Rightarrow \frac{|{S}|}{\bm{G}} \sim \frac{a}{r} \ll 1$,  so when the bead radius is small, the maximal value
 of elastic force is much smaller than gravity.
Fig. \ref{fig:forces} illustrates that in our simulations with $a \le 0.1$, the elastic forces are indeed much weaker than gravity.
For $a \ll r$, by neglecting the third term in \eqref{eq:dlugie}, we obtain the approximate dynamics which depends on the spring constant $k$ and the bead
radius $a$ only through the ratio $k/a$. The applicability of this approximation is consistent with our observations of trajectories with $r\sim 1$ in
the centre-of-mass reference frame for different values of $k$ and $a\le 0.1$, shown in Fig. \ref{fig:trajectories_fig}. 
The beads trajectories corresponding to the same ratio ${k}/{a}$ are almost the same.

\section{Discussion and concluding remarks}
In this work we report the novel finding of a horizontal hydrodynamic repulsion of two elastic dumbbells which settle down under gravity in a viscous fluid. The general remark following from our point-particle simulations  is that the replacement of rigid links between the dumbbells' beads by 
elastic springs significantly change the dynamics. 
The periodic tumbling, observed by \cite{Jung} for 
 two rigid sedimenting dumbbells,  
is replaced by a more complex pattern of non-periodic oscillations when 
 elastic dumbbells are initially located at the same configuration. In the limit of  vanishing spring constant $k$, when there is no constraints, we recover periodic tumbling, found by \cite{Hocking}.  The striking effects of elasticity are observed even though the elastic forces
 are very small in comparison to gravity (as demonstrated in section \ref{subsec:comparison}).

We have shown that for each value of $k$, there exists a universal time-dependent solution to which the system dynamics converges, as displayed in Fig. \ref{fig:ct}.
We have discovered the increase of the distance between the elastic dumbbells at their consecutive vertical 
configurations (shown in Fig.~\ref{fig:hyd_rep}), correlated with the systematic increase of the trajectory width (visible in Fig. \ref{fig:trajec}) at times corresponding to horizontal orientations of the dumbbells. The dumbbell length oscillates and very slowly tends to its equilibrium value. The maximal distance between dumbbells in vertical configurations is limited by the dumbbell length divided by the critical aspect ratio $C_0$; for $C<C_0$ tumbling motions do not occur. The hydrodynamic repulsion is faster for higher values of $C$, when the tumbling time is 
shorter. The hydrodynamic repulsion has been also observed for all values of $k$, with the largest speed for $k\!=\! 0.01-0.1$.  
The behaviour described above takes place for a wide range of values of $k$ and $a$, and in the lowest order, it depends only on the ratio~$k/a$ -- of course as long as the
point-particle approximation is justified.

The question arise what is the physical meaning of the dimensionless values of the spring constant, $k = 0.1$ and $k = 0.01$, for which the maximum repulsion effect is observed between sedimenting dumbbells. For example, let us consider springs made of DNA strands, which were used e.g. by \cite{Dreyfus} to construct artificial microswimmers. Within the Hookean regime, the spring constant of a DNA chain was measured e.g. by \cite{Bustamante}, with the corresponding values equal to 
$10^{-6} \frac{pN}{nm}$ - $10^{-5} \frac{pN}{nm}$ for
 decreasing values of the  chain length from $100 \mu m$ to $10 \mu m$. Constructing the dumbbells from $50\mu$m long strands of DNA, and beads twice as dense as water and radii around $5\mu$m, we recover the values of the dimensionless spring constant in the range $k=0.01-0.1$.

The mechanism of the observed convergence to the universal solution will be analysed
 elsewhere for a more general system of many elastic dumbbells.

\section*{Acknowledgement}
This work was supported in part by the Polish National Science Centre, 
Grant No. 2011/01/B/ST3/05691. Scientific benefits from the activities
of the COST Action MP1106 are acknowledged.

\end{document}